\documentclass{svmult}

\usepackage{makeidx}         
\usepackage{graphicx}        
\usepackage{multicol}        
\usepackage[bottom]{footmisc}
\usepackage{amsmath,amssymb}

\makeindex             

\begin{document}

\title*{Living Tissue Self-Regulation as\\ a Self-Organization Phenomenon}
\titlerunning{Living Tissue Self-Regulation Problem}
\author{Wassily Lubashevsky\inst{1} \and
        Ihor Lubashevsky   \inst{2} \and
        Reinhard Mahnke    \inst{3}
        }
  \institute{Moscow Technical University of Radioengineering, Electronics, and Automation,
    Vernadsky 78, 119454 Moscow, Russia \texttt{kloom@mail.ru}
\and
   A.M. Prokhorov General Physics Institute, Russian
    Academy of Sciences, Vavilov Str. 38, 119991 Moscow, Russia \texttt{ialub@fpl.gpi.ru}
\and
   Universit\"at Rostock, Institut f\"ur Physik, 18051 Rostock, Germany \texttt{reinhard.mahnke@uni-rostock.de}
}
\maketitle

\begin{abstract}
Self-regulation of living tissue as an example of self-organization phenomena in hierarchical systems of biological, ecological, and social nature is under consideration. The characteristic feature of these systems is the absence of any governing center and, thereby, their self-regulation is based on a cooperative interaction of all the elements. The work develops a mathematical theory of a vascular network response to local effects on scales of individual units of peripheral circulation.
\end{abstract}

\section{Introduction: Living Tissue Self-Regulation Problem}

Biochemical compounds (oxygen, nutrients, etc.) required by cells of human body for their life are transported by blood flow through the vascular network. Simultaneously, blood flow withdraws products of their life activity, e.g., carbon dioxide. So when cells in a certain region small in comparison with the whole organism consume an increased amount of oxygen and nutrients, blood flow through it just has to grow to maintain the internal environment of the cellular tissue. This effect will be referred to as the vascular network response to local variations in the state of the cellular tissue.

We consider the response of the vascular network to be \emph{perfect} if every cell is supplied with such amount of oxygen and nutrients that is strictly necessary for its current activity. As a result, if the functioning of a certain tissue region has changed and cells in it need an increased amount of oxygen and nutrients, then the vascular network with the perfect behavior is able to deliver the required biochemical compounds to this region in such a manner that blood flow at the other points of the cellular tissue not be disturbed at all.
However, in the general case, the vascular network responding to the local need of cells at one point will disturb the functioning of cellular tissue at all its other points. This effect can give rise to an erratic behavior of the organ as a whole and, thus, should be depressed. Since there are no special centers controlling blood flow redistribution on scales of individual organs this regulation can be implemented only via a cooperative interaction between all its elements. The latter feature allows us to call such regulation processes distributed self-regulation.

Up to now specific mechanisms by which the distributed self-regulation arises are far from being understood well. In order to construct an appropriate model for the distributed self-regulation two aspects must be elucidated. The first one is the self-processing of information because none of the blood vessels possesses the complete information about the state of the corresponding organ. Moreover, each vessel really needs only a piece of information required for its individual functioning. The second aspect concerns the cooperative mechanism governing the blood flow redistribution over the vascular network. Indeed, again, none of the vessels controls individually blood flow even through itself; responding to the appropriate information it can only dilate or constrict to change its hydrodynamic resistance to blood flow.

Here we present a model for the semi-perfect self-regulation of living tissue that does not require the vessels response to be ideal as was presumed previously \cite{we1,we2}, which enables us to describe observed phenomena in the real living tissue. The key point of the proposed model is the reaction of the blood vessels to the concentration of special biochemical components, activators, generated by the cells during their life activity most intensively when one of the internal environment parameters comes close to the tolerance boundary. The concept of activators can be justified appealing to biomedical data \cite{we3}.

\section{Living Tissue Model}\label{FM}


\begin{figure}
\begin{center}
\includegraphics[width=0.9\columnwidth]{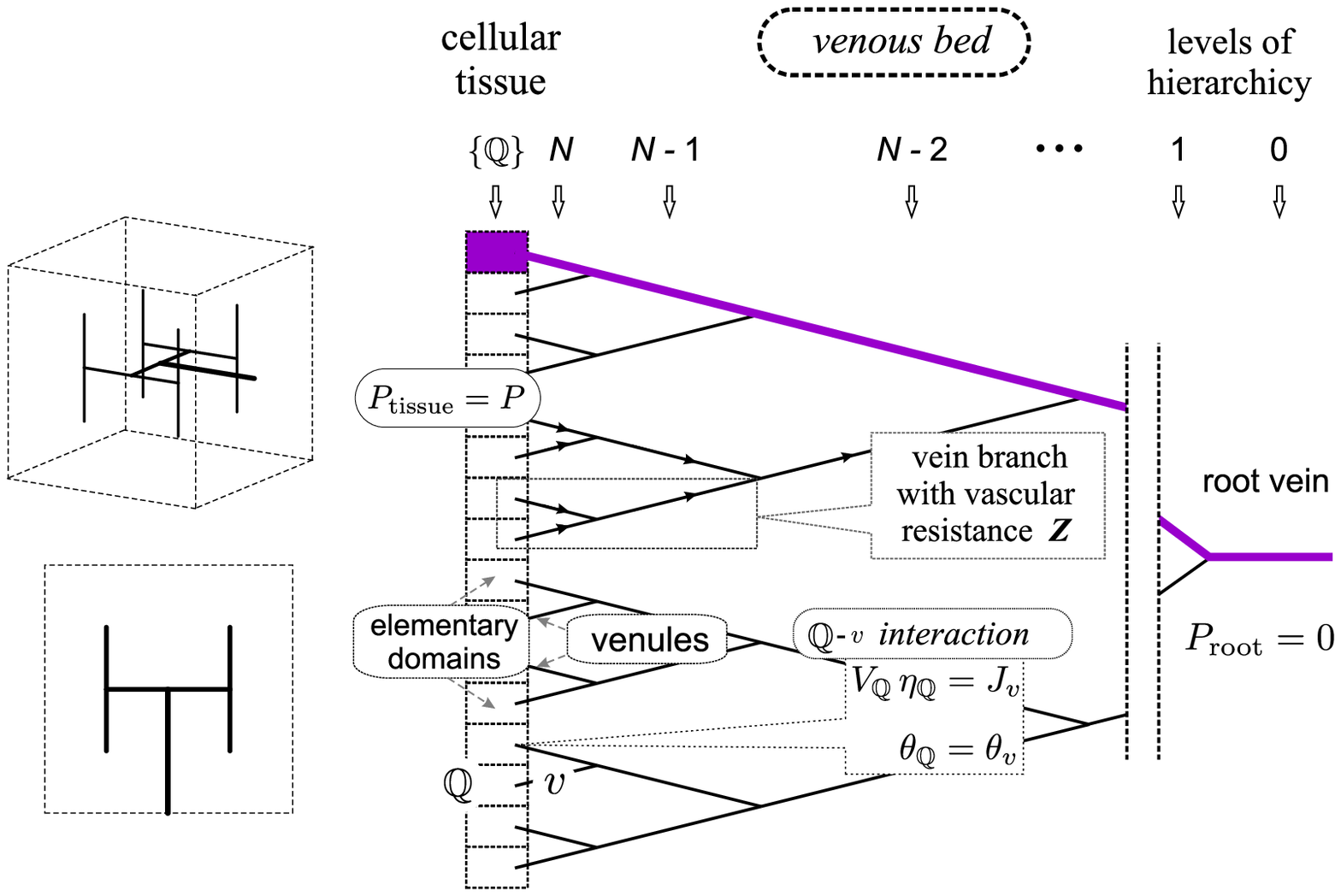}
\end{center}
\caption{Vascular network model reduced to the venous bed and its reaction to the distribution of activators in the cellular tissue. Here, in particular, $\mathcal{Q} = \mathcal{Q}_{c}$ designates the elementary domain related to the venule $c$. Left fragments illustrate the vessel embedding into the cellular tissue.}
\label{Fig1}
\end{figure}

The model for the vascular network architecture assumes the following. First, the arterial and venous beds are individually of the tree form and the branching of vessels is symmetrical. Second, the vessels of each level are distributed uniformly in the living tissue domain considered to be a square or cube. The first assumption enables us to introduce the vessel hierarchy and to order all the arteries and veins according to their position in the hierarchical structure. The second one allows us to specify the vessel arrangement using a self-similar embedding illustrated in Fig.~\ref{Fig1}. Besides, for the sake of simplicity, the arterial and venous beds are assumed to be the mirror image of each other with respect to all the properties, including the vessel response. This artificial assumption just simplifies the description of the vascular network response.


Blood flow distribution over the vein network $\{v\}$ (Fig.~\ref{Fig1}) is described as follows. First, to every vein $v$ we ascribe the blood flow rate $J_v$ and the activator concentration $\theta_v$ in blood going through it. The state of this vein is quantified in terms of the vessel resistance $R_n(\theta_v)$ depending on the activator concentration $\theta_v$, which reflects the vessel response. Here the function $R_n(\theta)$ is assumed to be identical for all the veins of one level $n$.  Second, to every branching point $k$ we ascribe a certain blood pressure $P$. Then the blood flow distribution is governed by the system of equations including Poiseuille's law written for every vein $v$
\begin{equation}\label{3.1}
    P_\text{in} - P_\text{out}  = J_v R_n(\theta_v)\,,
\end{equation}
and conservation of blood and activator substance at every branching point $k$
\begin{eqnarray}
\label{3.2}
    J_{\text{in},1} + J_{\text{in},2}  &=& J_{\text{out}}\,,
\\
\label{3.3}
    J_{\text{in},1}\theta_{\text{in},1} + J_{\text{in},2}\theta_{\text{in},2}
      &=& J_{\text{out}}\theta_{\text{out}}\,.
\end{eqnarray}
At the entrance to the venules $\{c\}$, i.e. the veins of the last level $N$ and the exit from the root vein (the vein of level $n= 0$)
\begin{equation}\label{3.4}
    P_{\text{in}|c}= P\,,\quad P_{\text{out}|n =0}=0\,.
\end{equation}
Finally, the equality of the activator concentration in every venule $c$ and the mean concentration of activators in the corresponding elementary domain $\mathcal{Q}_{c}$
\begin{equation}\label{3.5}
    \theta_{c}= \theta\left(\mathcal{Q}_{c}\right)
\end{equation}
holds. The elementary domains $\{\mathcal{Q}\}$ form the partition of the living tissue bulk in such way that every elementary domain contains just one venule, which is illustrated in Fig.~\ref{Fig1}. In the given analysis the distribution of activators in the cellular tissue specified in terms of $\{\theta(\mathcal{Q})\}$  is treated as given beforehand.

\begin{figure}
\begin{center}
\includegraphics[width=\columnwidth]{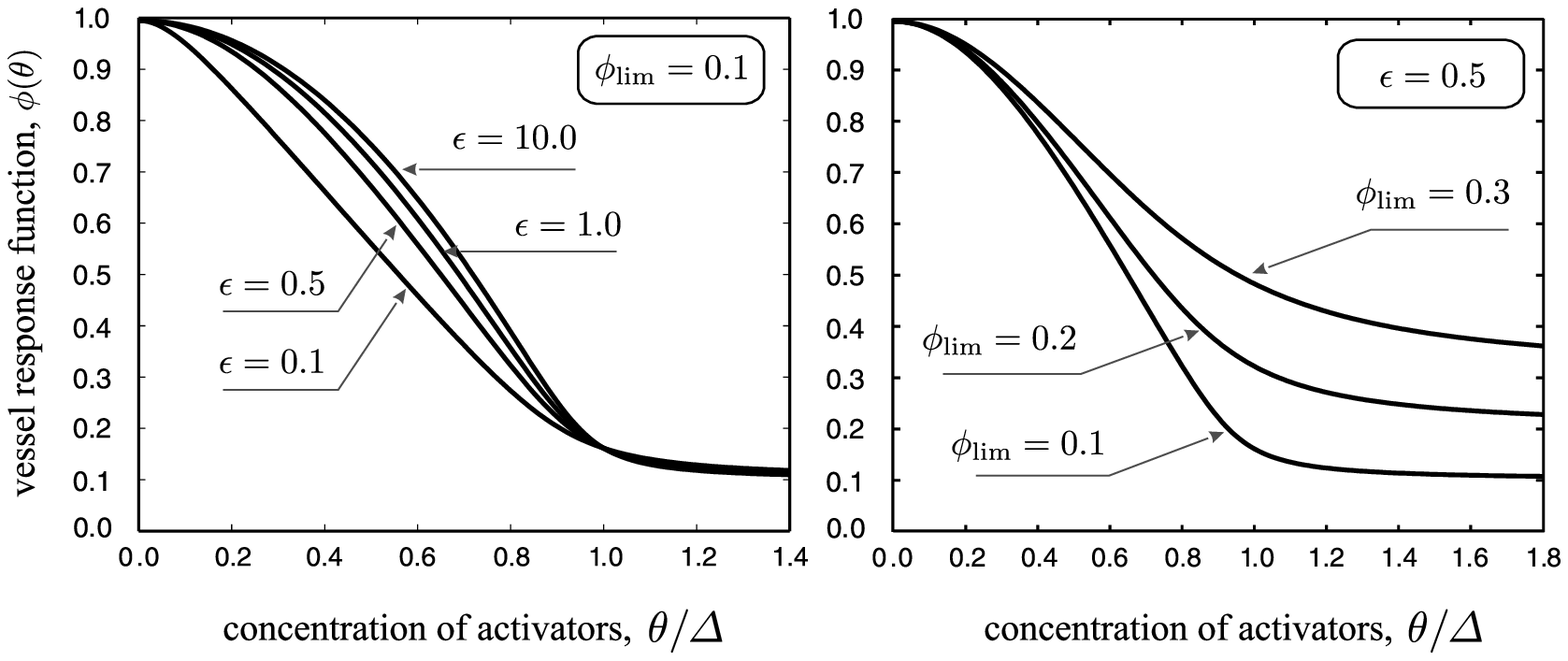}
\end{center}
\caption{The vessel response function $\phi(\theta)$ specified by ansatz~\eqref{3.10} for various values of its parameters.}
\label{Fig2}
\end{figure}

When one of the internal environment parameters comes close to the boundary of tolerance zone, the living tissue to survive has to increase essentially the blood flow rate via vessel dilation. So blood vessels should exhaust their ability to widen when the activator concentration exceeds a certain threshold $\Delta$ matching these critical conditions. It allows us to adopt the following ansatz
\begin{equation}\label{3.7}
    R_n(\theta_v) = \rho_n\phi(\theta_v)\,,
\end{equation}
where $\rho_n$ is the vessel resistance at $\theta_v = 0$ which depends only on the number of the hierarchy level and $\phi(\theta)$ is a certain function universal for all the vessels. By definition, at $\theta = 0$ the equality $\phi(0) =1$ holds and for $\theta\gg\Delta$ the inequality $\phi(\theta) \to \phi_\text{lim}\ll1$ should be the case. The following ansatz
\begin{align}
\label{3.10}
    \phi(\theta)& =\frac{\phi_\text{lim}+U(\theta)}{2}+\sqrt{\frac{[\phi_\text{lim}-U(\theta)]^{2}}{4}+2\phi_\text{lim}^{2}}
    \,,
\\
\intertext{with}
\label{3.11}
    U(\theta)&=\frac{1-\phi_\text{lim}-2\phi_\text{lim}^{2}}{1-\phi_\text{lim}}\cdot \frac{\sqrt{\epsilon^{2}+1}-\sqrt{\epsilon^{2}+(\theta/\Delta)^2}}{\sqrt{\epsilon^{2}+1}-\epsilon}\,.
\end{align}
is used to be specific. Here the additional numeric parameter $\epsilon$ characterizes the linearity of the function $\phi(\theta)$ in the region $\theta \lesssim \Delta$. Figure~\ref{Fig2} visualizes the given function $\phi(\theta)$ for various values of its parameters. The values $\epsilon = 0$ and $\phi_\text{lim} = 0$ match the ideal vessel response. Ansatz~\eqref{3.10} can be justified, for example, appealing directly to Fig.~\ref{Fig2}.

The dependence of the vessel resistance on the hierarchy level, i.e. the value of $\rho_n$ vs the level number $n$ is constructed using the known equality of contributions to the vascular resistance of arterial beds from all the hierarchy levels. This enables us to write the $\rho_n$-dependence in the form
\begin{equation}\label{3.12}
    \rho_n=(2\zeta)^{n}\rho_0\,,
\end{equation}
where the value $\rho_0$ is related with the root vein and the parameter $\zeta\approx1$ allows us to consider the cases where various groups of vessels dominate in the blood pressure distribution. In particular, for $\zeta < 1$ the large veins contribute mainly to the vascular resistance of the venous bed, whereas, in the opposite case, $\zeta > 1$, the small veins are dominating.

\section{Results}\label{Sec:Res}

\begin{figure}
\begin{center}
\includegraphics[width=\columnwidth]{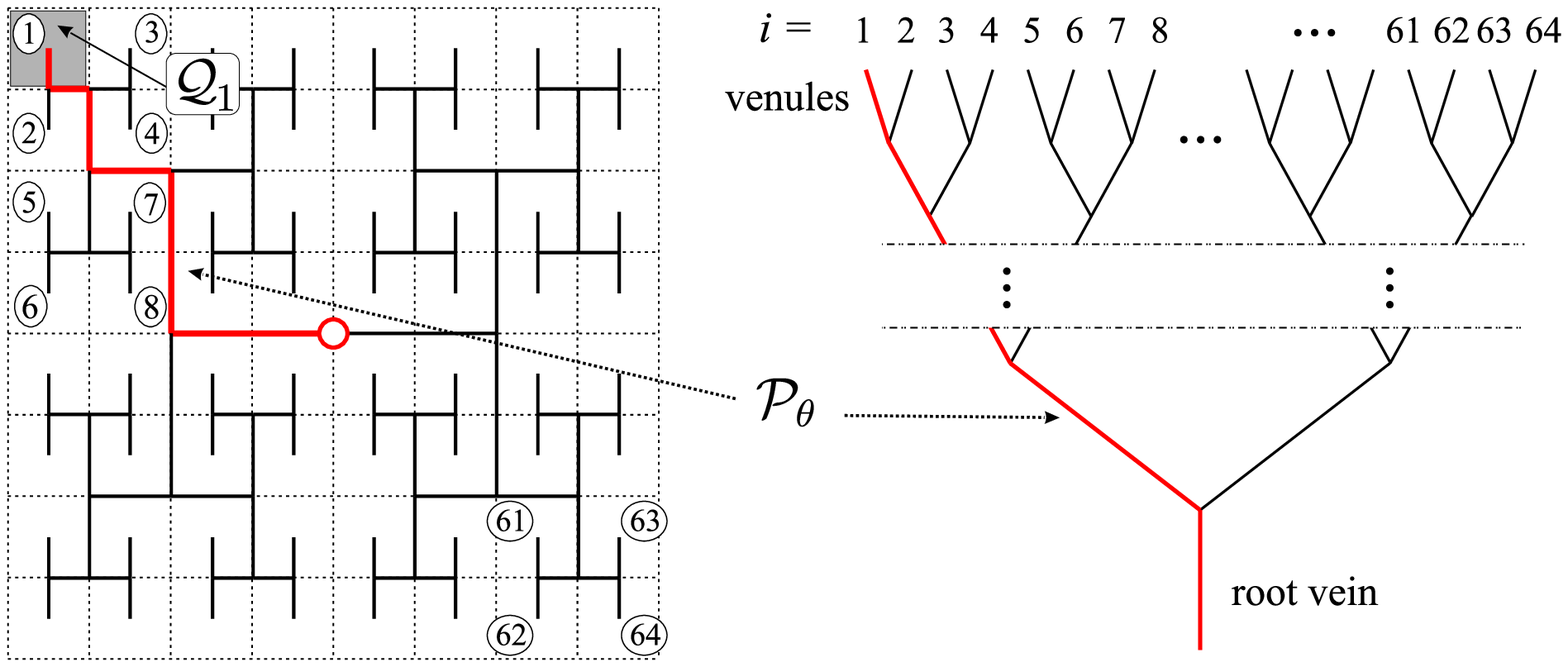}
\end{center}
\caption{Example of the systems analyzed numerically. The region with a nonzero value of the activator concentration $\theta$ is shadowed and the ordering of the venules is illustrated for the system with the number of the hierarchy levels equal to $N=6$.}
\label{Fig3}
\end{figure}

The blood flow redistribution governed by equations~\eqref{3.1}--\eqref{3.3} subjected to conditions~\eqref{3.4} and \eqref{3.5} was studied numerically. Let us, first, present the results for the case when the activator concentration $\theta$ differs from zero inside $m$ first elementary domains located at one of the square corner (2D model of living tissue, Fig.~\ref{Fig3}), namely,
\begin{align}\label{5.1}
 \theta_{c,i} = \theta &&\text{if\quad $1\leq i \leq m$} && \text{and}&& \theta_{c,i} = 0 && \text{if\quad $1< m \leq N$} \,.
\end{align}
Figure~\ref{Fig3} illustrates the analyzed case and shows the introduced ordering of the venules with index $i$. The elementary domains are also labeled with the index $i$ of the corresponding venule.

\begin{figure}
\begin{center}
\includegraphics[width=\columnwidth]{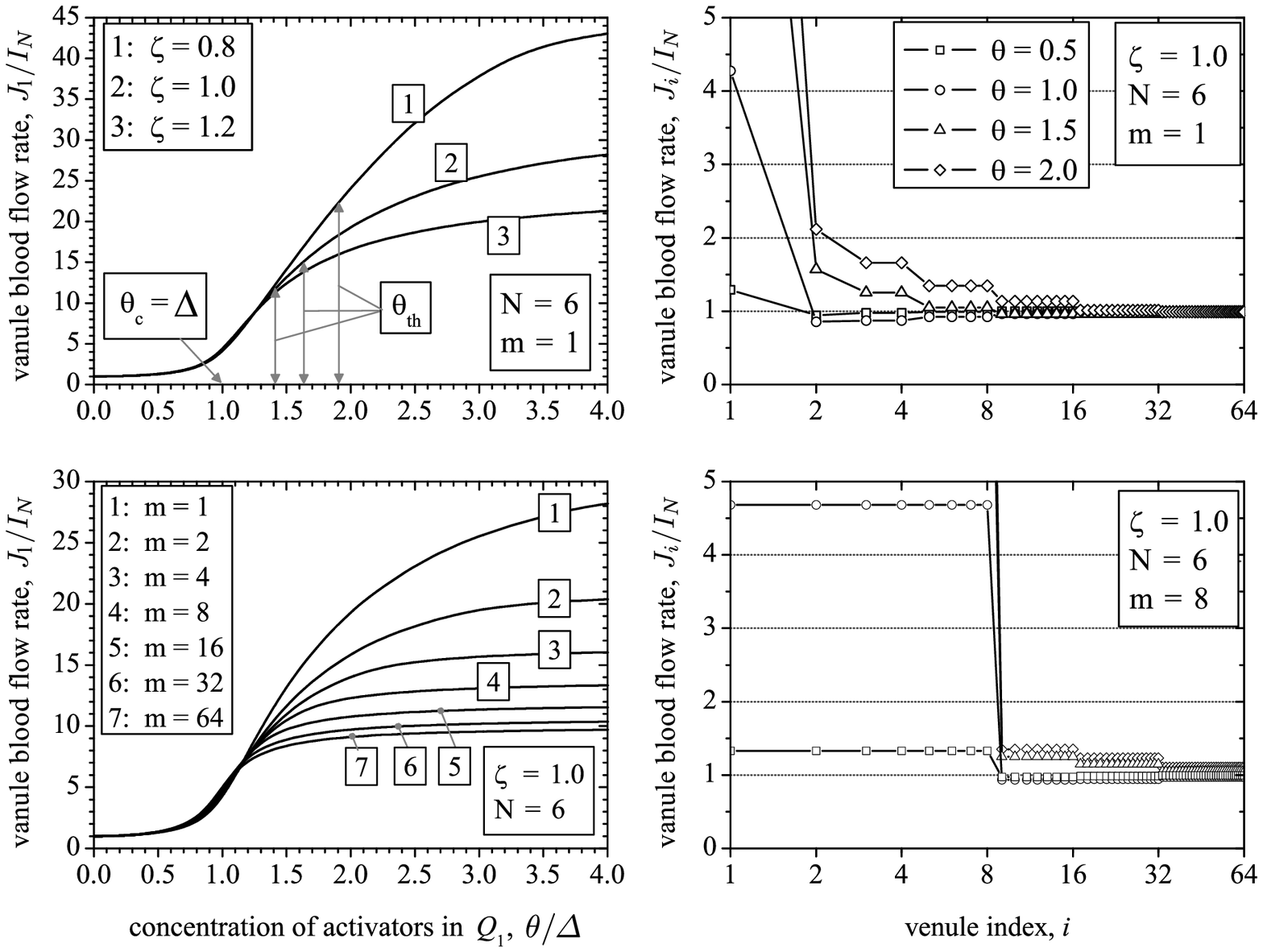}
\end{center}
\caption{Results of numerical analysis. The blood flow rate  $J_1$ (in units of $I_N$) vs the concentration of activators in the elementary domain $\mathcal{Q}_1$ (left column) and the distribution of the blood flow rates (in units of $I_{N}$) over the venules (right column). In obtaining these results the parameters $\epsilon = 1.0$ and $\phi_\text{lim} = 0.1$ were used, the used values of the other parameters are shown in figure.}
\label{Fig4}
\end{figure}

Figure~\ref{Fig4} presents the obtained results. The first column exhibits the dependence of the blood flow rate $J_1$ through the venule $v_1$ on the activator concentration $\theta$ in units of the blood flow rate $I_N$ through venules under the normal conditions, i.e. in the case of absence of activators. The number of the hierarchy levels was set equal to $N=6$; it meets tenfold decrease in the vessels length when passing from the root vessel to the smallest arterioles or venules.

The left upper frame depicts three curves corresponding to different values of the parameter $\zeta$. When the activator concentration in the domain $\mathcal{Q}_1$ gets the threshold $\theta_c = \Delta$ of the individual vessel reaction the venule joined with this domain  exhausts its capacity to widen. However it is not enough for the blood flow rate to grow substantially. Indeed, the contribution of this venule to the vascular resistance is small even along the path $\mathcal{P}_\theta$ leading from the root vessel to the domain $\mathcal{Q}_1$ on the vein tree (Fig.~\ref{Fig3}). So many veins along this path have to dilate in order for the blood flow rate to grow essentially in $\mathcal{Q}_1$. However, the larger a vein, the larger the tissue region drained through this vein as a whole and its size increases exponentially with the number of branching points separating the venule $v_1$ and the given vein along $\mathcal{P}_\theta$. As a result when the activator concentration in the domain $\mathcal{Q}_1$ is about $\theta \sim \Delta$ the activator concentration in relatively large veins on the path $\mathcal{P}_\theta$ turns out to be rather small and these vessels cannot widen remarkably. Thereby the blood flow rate cannot also exhibit substantial increase (Fig.~\ref{Fig4}). As the activator concentration grows further the induced increase in the activator concentration along the path $\mathcal{P}_\theta$ gets the critical value $\Delta$ in larger and larger veins, causing the growth of the blood flow rate $J_1$. When the main part of the veins belonging to $\mathcal{P}_\theta$ widen to the maximum the blood flow rate comes to the upper limit $J_\text{max}$.

The maximum $J_\text{max}$ of the blood flow rate depends on the system parameter, in particular, on $\zeta$. As it will be demonstrated below practically all the additional amount of blood entering the system due to the vessel dilation is directed to the domain $\mathcal{Q}_1$. So the higher the contribution of large veins, the higher the upper limit of the blood flow rate. These speculations are justified well in Fig.~\ref{Fig4}, the left upper frame. Because of this blood flow focusing along the path $\mathcal{P}_\theta$ the value $J_\text{max}$ can exceed substantially the value $1/\phi_\text{lim}$ which would be attained if the total blood pressure drop had fallen just on the last vessel supplying the region $\mathcal{Q}_1$ with blood. A similar effect should be expected when the region of living tissue where activators are located becomes larger. So the value of $J_\text{max}$ has to drop with $m$ increasing, which is demonstrated in Fig.~\ref{Fig4} (left bottom frame). The value $J_\text{max}$ goes down to $\phi_\text{lim}$ when activators spread over the whole region of the microcirculatory bed.

The obtained results demonstrate us also the fact that when the vessel behavior is not ideal the threshold $\theta_\text{th}$ in the vascular network response as a whole does not coincide with the critical value $\Delta$ of the individual vessel reaction and can exceed it remarkably. Let us treat the threshold $\theta_\text{th}$ as the value of $\theta$ at which the blood flow rate attains one half of its upper limit. Then this fact is clearly visible in Fig.~\ref{Fig4} (left upper frame) and, for example, the estimate $\theta_\text{th}\approx 2\Delta$ holds for the case shown by curve~1. Naturally, the threshold $\theta_\text{th}$ depends also on the parameters of the vascular network but not only on the characteristics of individual vessel behavior.

The shown curves enable us to assume that due to the hierarchical structure of the vascular network there is a certain function $\eta(\theta|\phi,\epsilon):=J/I_N$ depending only on the parameters of individual vessel behavior, which describes the increase in the blood flow rate with the growth of the activator concentration until the latter becomes greater then the threshold $\theta_\text{th}$. The other characteristics of the vascular network and the distribution of activators over the cellular tissue, roughly speaking, affect only the upper limit $J_\text{max}$ of blood flow rate.

The right column depicts the distribution of the blood flow rates over the venules. Comparing the data shown hear and that of the left column we can declare that the main increase in the blood flow rate is located along the path $\mathcal{P}_\theta$. However, the question about the interference of blood streams induced by distant regions of living tissue requires an individual investigation. Nevertheless, when the living tissue is excited only in one domain its response can be treated as quasilocal. We have used the term ``quasilocal'' because this response also depends on the size of the excited domain and contains a certain component describing small variations in the blood flow rate at distant points.

As the nonlocal component of the $J_i\{\theta\}$-functional is concerned, the right column of Fig.~\ref{Fig4} demonstrates us that depending on the relation between the activator concentration $\theta$ in the elementary domain $\mathcal{Q}_1$ and the vascular network threshold $\theta_\text{th}$ it changes its form. When $\theta< \theta_\text{th}$ the increase in the blood flow rate through $\mathcal{Q}_1$ is caused by the blood flow decrease through the other elementary domains. In this case the required growth of the blood flow rate is due to  the redistributed of blood flow over the vascular network without additional amount of blood entering the microcirculatory bed. Since the number of the surrounding domains is rather large the decrease in the blood flow rate outside the excited region is not essential. When $\theta> \theta_\text{th}$ the increase in the blood flow rate is mainly caused by additional portion of blood entering the microcirculatory bed. Therefore the blood flow rate is increased in the surrounding domains also.

\begin{figure}
\begin{center}
\includegraphics[width=\columnwidth]{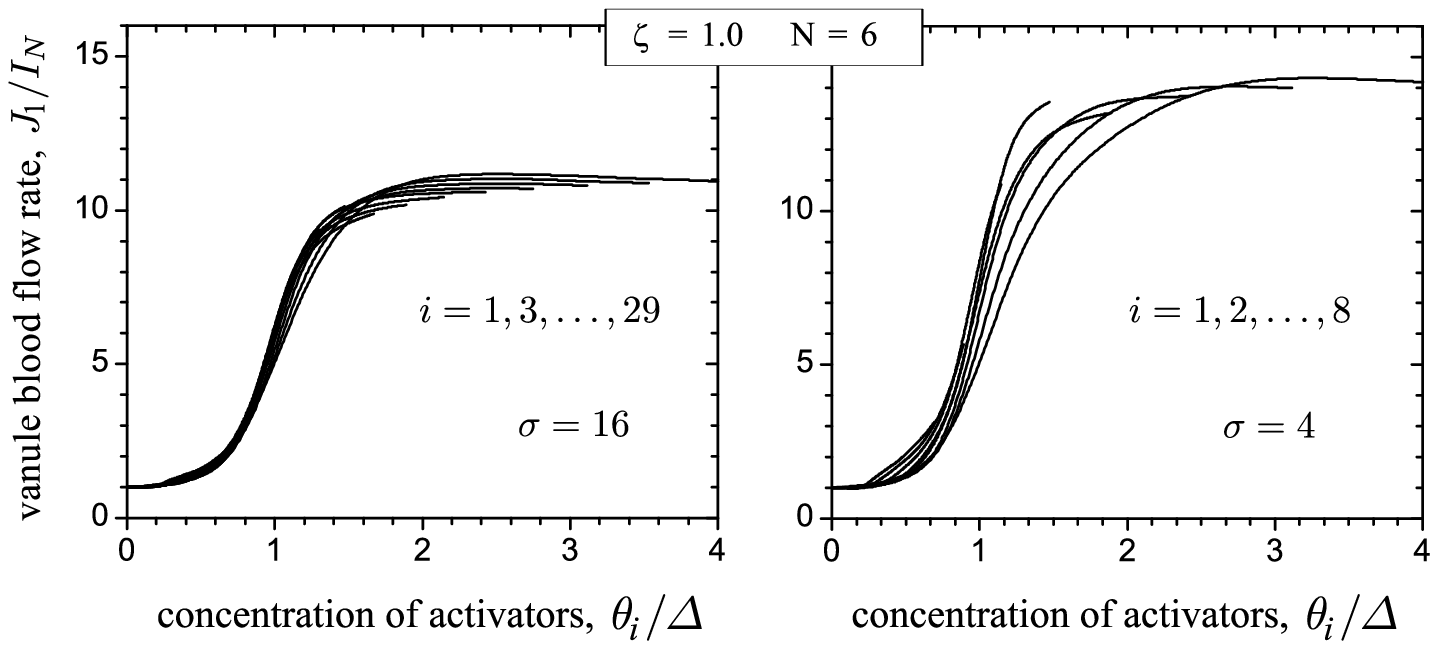}
\end{center}
\caption{Results of numerical analysis. The family of the dependencies $\{J_i(\theta_{i})\}$ each of them is the rate $J_i$ of blood flow through venule $i$ vs the activator concentration $\theta_i$ in it for the activator distribution given by expression~\eqref{5.2}. In obtaining these results the parameters $\epsilon = 1.0$ and $\phi_\text{lim} = 0.1$ were used, the used values of the other parameters are shown in figure.}
\label{Fig5}
\end{figure}

To analyze the validity of the quasi-locality approximation of the living tissue response we have studied the blood flow rate distribution $J_i$ over the venules for the exponential form of the activator distribution over the cellular tissue, namely,
\begin{align}\label{5.2}
 \theta_{c,i} = \theta \exp\left[-\frac{(i-1)}{\sigma}\right]
\end{align}
for different values of the parameter $\sigma$. Figure~\ref{Fig5} exhibits the resulting dependence of the rate $J_i(\theta_i)$ of blood flow through venule $i$ on the local value $\theta_i$ of the activator concentration in this venule. Plotting the family of the curves $J_i(\theta_i)$ for a collection of venules we can verify whether the dependence $J_i(\theta_i)$ is really a local function of its argument. The found results (Fig.~\ref{Fig5}) demonstrate us that when the activators are concentrated in the region whose size is larger then the size of elementary domain the blood flow rate $J_i$ does depend only on the local value of the activator concentration $\theta_i$. Only when the region of activator localization is comparable with the elementary domains in size the quasi-locality approximation is violated to some degree.

\section{Conclusion}

We have presented a model for the distributed self-regulation of living tissue that exemplifies a general mechanism by which self-regulation of active hierarchical systems without controlling centers can arise. The key point is the information self-processing caused by conservation of blood and activators in blood stream through the arterial and venous beds. It is essential that the vascular network response is based on vessels reacting to certain biochemical components, activators, that are generated by the cells when one of the internal environment parameters comes close to the tolerance boundary. In this case under rather general assumptions about the specific properties of this reaction the vascular network response to local perturbations is quasi-local. In other words the quasi-locality is the consequence of the hierarchical structure of the vascular network rather then the ideal behavior of its elements as though previously \cite{we1,we2}. In this way we have overcome the main obstacle to applying the developed theory of distributed self-regulation to describing real phenomena in such systems. Besides, due to the hierarchical structure of the vascular network its response turns out to be much stronger then it could be expected considering the individual dilation of blood vessels.   
     
\vspace*{\baselineskip}
\emph{Acknowledgments:} The authors appreciates the support of DFG Grant MA 1508/8-1, RFBR Grants 06-01-04005 and 09-01-00736 as well as the research support R-24-4 from the University of Aizu.



\printindex
\end{document}